\documentclass{article}
\usepackage{amsmath}
\usepackage{amssymb}

\input{tcilatex}

\begin{document}

\begin{center}
\textbf{\ Unified Statistical Description of Quasithermodynamic Systems in
and out of Equilibrium }

\bigskip

E.D. Vol

\bigskip

\textit{B. Verkin Institute for Low Temperature Physica and Engineering of
the National Academy of Sciences of Ukraine, 47 Lenin ave., 61103 Kharkov,
Ukraine}

Abstract
\end{center}

We propose the method of statistical description of broad class of dynamic
systems (DS) whose equations of motion are determined by two state depending
functions: 1) "energy" - the quantity which conserves in time and 2)
"entropy" - the quantity which does not decrease in time. It is demonstrated
that the behavior of such systems in the equilibrium state reduces to the
thermodynamic lows in particular the Le Chatelier principle is satisfied and
so on. Taking into account the interaction system of interest with ergometer
-- the device which continuously measures its energy one can possible to
find the system distribution function in arbitrary non-equilibrium
stationary state (NESS). Some general relations for mean values of certain
quantities in NESS which can be compared with experimental data are obtained.

\newpage

\section{Introduction}

The dynamic equations method is the base mathematical tool for investigation
of the behavior of complex systems. Such method of description is widely
used not only in classical physics where it appeared for the first time but
also in chemistry, biology, ecology and other sciences. The types of
equations describing sufficiently broad classes of systems (and the
processes taking place in them) on the one hand and having some general
features which do not depend on the nature of the studying objects on the
other hand are especially important. In this connection we should point out
such classes of equations:1) Hamiltonian ones and 2) equations of gradient
type. Hamiltonian equations of motion have the form: $\frac{dx_{i}}{dt}=%
\frac{\partial H}{\partial p_{i}};\frac{dp_{i}}{dt}=-\frac{\partial H}{%
\partial x_{i}}$, where $H(x_{i},p_{i})$ is the Hamilton function of the
system. They are used for describing the evolution of conservative DS with
even-dimensional phase space whose behavior is reversible in time. For
describing the nonreversible systems with the strong dissipation the
equations of gradient type are usually used: $\frac{dx_{i}}{dt}=-\frac{%
\partial \Phi }{\partial x_{i}}$,where $\Phi (x_{i})$ - Lyapunov function of
the system, the quantity which monotonically decreases in time.

Let us now call attention to the study of evolution of closed system with
the large number of degrees of freedom. It is well known that such system
comes to the equilibrium state without fail at the condition $t\rightarrow
\infty $ (zero law of thermodynamics). In his seminal paper of 1865, R.
Clausius stated the laws determining the behavior of such systems in the
form of the following two propositions(see \cite{1},p. 194):

\begin{equation}
\text{1) Die Energie der Welt ist constant}
\end{equation}

\begin{equation*}
\text{2) Die Entropie der Welt strebt einem Maximum zu}
\end{equation*}

We'll consider the class of DS which satisfy to this Clausius conditions. In
this connection we want to ask two questions: 1) What equations of motion
with necessity provide such behavior of the system? 2) What results related
to the statistical behavior of such systems in and out of equilibrium can be
obtained from such equations of motion? The attempt to answer these
questions is the main goal of our paper. Note that in phenomenological
approach which is used in this paper the terms "energy" and "entropy" will
be used only as conventional labels for two quantities satisfying the
Clausius conditions. Now we describe briefly the content of the paper. It
consists of five sections. In Sec.1 we introduce the notion of
quasithermodynamic system (QTS) and give the concrete recipes to construct
the equations of motion for such systems. In Sec.2 on the base of these
equations we investigate the conditions of equilibrium of two QTS. We prove
that in the case when an interaction between two QTS is absent our
conditions of equilibrium are reduced to the standard thermodynamic
relation: $\frac{\partial S_{1}}{\partial E_{1}}=\frac{\partial S_{2}}{%
\partial E_{2}}$. Besides using the condition of equilibrium state stability
the Le Chatelier principle for QTS is derived directly from equations of
motion. In Sec.3 we propose the method of statistical description of QTS
behavior. In order to avoid the appearance of singular distribution
functions we introduce the interaction between the system of interest and
the ergometer- device which continuously measures its energy. Such method
gives us the possibility to formulate the simple Fokker-Planck equation for
QTS distribution function whose stationary solution can be obtained. This
solution can be expressed via state functions $S(x_{i})$ and $E(x_{i})$ and
additional parameter $\beta $. In the limiting case of equilibrium state the
solution obtained reduces to the ordinary Gibbs distribution: $f(x_{i})%
\symbol{126}\exp (-\beta E(x_{i}))$. In Sec.4 we investigate the
non-equilibrium stationary states of QTS more detail and formulate
some general relations for them. In the last part of the paper we
summarize the results and briefly discuss the possible
applications of them. Now let us pass to the detailed account.

\section{Section 1}

We define as quasitermodynamic (of N order) the DS whose state is described
by $N$ variables $x_{1,}..,x_{N}$ and whose behavior in time satisfies two
Clausius conditions (1). Meanwhile we suppose that variables $x_{1,}..,x_{N}$
unlike from coordinates and momenta define the "coarse -grained" description
of macroscopic system evolution and therefore the number of them can be
arbitrary. Let us establish now the form of equations of motion for given
state functions: energy $H(x_{1,}..,x_{N})$ and the entropy $%
S(x_{1,}..,x_{N})$ providing the realization of the distinctive properties
of QTS :

\begin{equation}
\frac{dH}{dt}=\frac{dH}{dx_{i}}\overset{\cdot }{x}_{i}=0,  \label{1a}
\end{equation}

\begin{equation}
\frac{dS}{dt}=\frac{dS}{dx_{i}}\overset{\cdot }{x}_{i}\geqslant 0.
\label{1b}
\end{equation}

(In relations (2), (3) and later on we use the usual agreement about the
summation on repeating indexes).

We start from the simplest case of QTS of second order ($N=2$) and suppose
that required system of equations of motion is an autonomous system of
differential equations: $\frac{dx_{i}}{dt}=\Phi _{i}(x_{1},x_{2})$. By
direct verification one can be convinced that taking functions $\Phi
_{i}(x_{1},x_{2})$ in the form: $\Phi _{i}\equiv \varepsilon _{ik}\frac{%
\partial H}{\partial x_{k}}\left\{ S,H\right\} $, where $\varepsilon
_{ik}\equiv \left(
\begin{array}{cc}
0 & 1 \\
-1 & 0%
\end{array}%
\right) $ is antisymmetric tensor of 2 rank and $\{S,H\}$ is Poisson bracket
of functions $S(x_{1},x_{2})$ and $H(x_{1},x_{2})$, i.e. by the definition $%
\left\{ S,H\right\} =\frac{\partial S}{\partial x_{1}}\frac{\partial H}{%
\partial x_{2}}-\frac{\partial S}{\partial x_{2}}\frac{\partial H}{\partial
x_{1}}$, we obtain the equations of motion

\begin{equation}
\frac{dx_{i}}{dt}=\varepsilon _{ik}\frac{\partial H}{\partial x_{k}}\left\{
S,H\right\} ,  \label{2}
\end{equation}

from which one can easily obtain properties (2) and (3).

Similarly in the case $N=3$ one can be convinced that the system of
equations of motion with required properties can be presented in the form:

\begin{equation}
\frac{dx_{i}}{dt}=\varepsilon _{ikl}\frac{\partial H}{\partial x_{k}}A_{l},
\label{3}
\end{equation}

where $\varepsilon _{ikl}$ is the antisymmetric tensor of 3 rank and

\begin{equation}
A_{l}\equiv \varepsilon _{lmn}\frac{\partial S}{\partial x_{m}}\frac{%
\partial H}{\partial x_{n}}.  \label{3a}
\end{equation}

In addition we give the form of equations of motion for the case $N=4$:

\begin{equation}
\frac{dx_{i}}{dt}=\varepsilon _{iklm}\frac{\partial H}{\partial x_{k}}A_{lm},
\label{4}
\end{equation}

where $\varepsilon _{iklm}$ is the antisymmetric tensor of 4 rank and

\begin{equation}
A_{lm}\equiv \varepsilon _{lmnp}\frac{\partial S}{\partial x_{n}}\frac{%
\partial H}{\partial x_{p}}.  \label{4a}
\end{equation}

The generalization of equations of motion in the case of arbitrary $N$ is
obvious. We will not discuss here the rigorous mathematical justification of
the uniqueness of equations obtained. About this problem one can see the
paper \cite{2} and references in it. In this paper we are interested in the
properties of QTS implied from equations of motion (4), (5), (7) and the
possibility of statistical description of such systems.

\section{Section 2}

We start our analysis of equations of motion for QTS from the simplest case $%
N=2$.

Let us rewrite the equations (2) into another, more convenient form. For
this purpose we introduce the "phase' $\varphi $ which is variable
canonically conjugated to the energy i.e. $\{\varphi ,H\}=1$. Now one can
easily see that the system (2) is equivalent to the following system of
equations for H and $\varphi $:

\begin{equation}
\frac{dH}{dt}=0,\frac{d\varphi }{dt}=\frac{\partial S}{\partial \varphi },
\label{5}
\end{equation}

where QTS entropy $S$ now must be expressed via $\varphi $ and $H$. It
follows from (9) that in the case when $S(\varphi ,H)$ has maximum on $%
\varphi $ there is the line i.e. the one-parameter set of equilibrium points
$\varphi =\varphi _{0}(E)$ corresponding to the solution of equation $\frac{%
\partial S(\varphi ,E)}{\partial \varphi }=0$ when $\varphi =\varphi _{0}(E)$%
. Similar equilibrium condition is valid in general case also. It is easy to
show that equilibrium conditions for QTS of $N$-order reduce to the next
system of equations (it follows from equality to zero the right parts of
(4), (5), (7)):

\begin{equation}
\frac{\frac{\partial S}{\partial x_{1}}}{\frac{\partial H}{\partial x_{1}}}=%
\frac{\frac{\partial S}{\partial x_{2}}}{\frac{\partial H}{\partial x_{2}}}%
=...\frac{\frac{\partial S}{\partial x_{N}}}{\frac{\partial H}{\partial x_{N}%
}}\equiv \lambda .  \label{6}
\end{equation}

Now let us consider the composite system consisting of two subsystems: A
with variables $x_{1,}..,x_{N1}$ and B with coordinates $y_{1,}..,y_{N2}$.
Besides we suppose that subsystems A and B are non-interacting i.e.
relations $H=E_{A}(x_{1,}..,x_{N1})+E_{B}(y_{1,}..,y_{N2})$ and $%
S=S_{A}(x_{1},..,x_{N1})+S_{B}(y_{1,}..,y_{N2})$. In this case equilibrium
conditions reduce to the form:

\begin{equation}
\frac{\frac{\partial S_{A}}{\partial x_{1}}}{\frac{\partial H_{A}}{\partial
x_{1}}}=...\frac{\frac{\partial S_{A}}{\partial x_{N1}}}{\frac{\partial H_{A}%
}{\partial x_{N1}}}=\frac{\frac{\partial S_{B}}{\partial y_{N1}}}{\frac{%
\partial H_{B}}{\partial y_{N1}}}...\frac{\frac{\partial S_{B}}{\partial
x_{N2}}}{\frac{\partial H_{B}}{\partial x_{N2}}}.  \label{7}
\end{equation}

Equations (11) can be satisfied by setting $S_{A}=S_{A}(E_{A})$ and $%
S_{B}=S_{B}(E_{B})$ after that equations reduce to standard condition of
thermodynamic equilibrium of two subsystems:

\begin{equation}
\frac{\partial S_{A}}{\partial E_{A}}=\frac{\partial S_{B}}{\partial E_{B}}.
\label{8}
\end{equation}

Now we demonstrate how using only equilibrium equations (10) one can deduce
the Le Chatelier principle. We remind that Le Chatelier principle claims
that response of the system in the equilibrium state on the external
influence always results in the decreasing of this influence. For proving
this principle in the simplest case of 2-order QTS with variables $x$ and $y$
let us consider three states of such system: I - ($x_{0},y_{0}$) - the
initial equilibrium state, II - ($x_{0},y_{0}+\delta y$) - nonequilibrium
state enforced by small influence on the variable y, III - ($x_{0}+%
\widetilde{\delta x},y_{0}+\widetilde{\delta y}$) - the final equilibrium
state. It is obvious that the final equilibrium state III differs from
initial equilibrium state I. Because the energy of the system have been
changed. The first condition we write down is equality of energies of states
II and III: $H_{II}=H_{III}=H(x_{0},y_{0})+\delta H$, where

\begin{equation}
\delta H=\left( \frac{\partial H}{\partial y}\right) _{0}\delta y=\left(
\frac{\partial H}{\partial x}\right) _{0}\widetilde{\delta x}+\left( \frac{%
\partial H}{\partial y}\right) _{0}\widetilde{\delta y}.  \label{9}
\end{equation}

Besides both states I and III are equilibrium states the following
conditions must be satisfied:

\begin{equation}
\frac{\left( \frac{\partial S}{\partial x}\right) _{0}}{\left( \frac{%
\partial H}{\partial x}\right) _{0}}=\frac{\left( \frac{\partial S}{\partial
y}\right) _{0}}{\left( \frac{\partial H}{\partial y}\right) _{0}}\equiv
\lambda  \label{10a}
\end{equation}

and

\begin{equation}
\frac{S_{x}+S_{xx}\widetilde{\delta x}+S_{xy}\widetilde{\delta y}}{%
H_{x}+H_{xx}\widetilde{\delta x}+H_{xy}\widetilde{\delta y}}=\frac{%
S_{y}+S_{yx}\widetilde{\delta x}+S_{yy}\widetilde{\delta y}}{H_{y}+H_{yx}%
\widetilde{\delta x}+H_{yy}\widetilde{\delta y}},  \label{10b}
\end{equation}

where for simplicity we used notations: $S_{x}\equiv \left( \frac{\partial S%
}{\partial x}\right) _{0}=\left( \frac{\partial S}{\partial x}\right)
_{\left( x_{0},y_{0}\right) }$; $S_{y}\equiv \left( \frac{\partial S}{%
\partial y}\right) _{0}$ and so on. We can write $\widetilde{\delta x}%
=A\delta y$ and $\widetilde{\delta y}=B\delta y$, where $A$ and $B$ are yet
unknown functions of $x_{0}$ and $y_{0}$. Equations (13) and (15) now are
reduced to the form:

\begin{equation}
H_{y}=H_{y}B+H_{x}A,  \label{11a}
\end{equation}

\begin{equation}
\frac{\left( S_{xx}A+S_{xy}B\right) H_{x}-S_{x}\left( H_{xx}A+H_{xy}B\right)
}{H_{x}^{2}}=  \label{11b}
\end{equation}

\begin{equation*}
=\frac{\left( S_{yx}A+S_{yy}B\right) H_{y}-S_{y}\left(
H_{yx}A+H_{yy}B\right) }{H_{y}^{2}}.
\end{equation*}

We can simplify the form of equations (16) and (17) by using the quantity $%
F=S-\lambda H$. Then taking into account (14) for unknown functions $A$ and $%
B$ we obtain the following system

\begin{equation}
H_{x}A+H_{y}B=H_{y},\qquad \left( \frac{F_{xx}}{H_{x}}-\frac{F_{xy}}{H_{y}}%
\right) A+\left( \frac{F_{xy}}{H_{x}}-\frac{F_{yy}}{H_{y}}\right) B=0.
\label{12}
\end{equation}

Solving the system (18) we obtain

\begin{equation}
A=\frac{H_{y}}{\Delta }\left( H_{x}F_{yy}-H_{y}F_{xy}\right) ,\qquad B=\frac{%
H_{y}}{\Delta }\left( H_{y}F_{xx}-H_{x}F_{xy}\right) ,  \label{13}
\end{equation}

where $\Delta \equiv H_{y}^{2}F_{xx}-2H_{y}H_{x}F_{xy}+H_{x}^{2}F_{yy}$.

Note that equilibrium stability conditions for state I require inequalities $%
F_{xx}<0$, $F_{yy}<0$ and $F_{xx}F_{yy}-F_{xy}^{2}>0$ are valid. In
particular this case follows that $\Delta <0$.

Now for proving the Le Chatelier principle it is necessary to introduce the
measure of QTS reaction on the exterior influence. As such measure it is
natural to choose the magnitude of thermodynamic force $Y\equiv -\left(
\frac{\partial F}{\partial y}\right) $ that is equal to zero in unperturbed
state I. Then for state II we have following relation: $\Delta Y\equiv
Y_{II}-Y_{I}=\left( \frac{\partial Y}{\partial y}\right) _{0}\delta
y=-F_{yy}\delta y$ and for equilibrium state III $\Delta \widetilde{Y}\equiv
Y_{III}-Y_{I}=\left( \frac{\partial Y}{\partial x}\right) _{0}\widetilde{%
\delta x}+\left( \frac{\partial Y}{\partial y}\right) _{0}\widetilde{\delta y%
}=-\left( F_{xy}A+F_{yy}B\right) \delta y$.

The Le Chatelier principle will be valid under the condition

\begin{equation}
F_{xy}A+F_{yy}B\geqslant F_{yy}.  \label{14}
\end{equation}

Substituting $A$ and $B$ from (19) into (20) after some simple algebra we
came to the inequality $F_{yy}^{2}H_{x}^{2}+H_{y}^{2}F_{xy}^{2}\geqslant
2H_{x}H_{y}F_{xy}F_{yy}$, which is obviously valid. Thus the validity of the
Le Chatelier principle is proved.

\bigskip

\section{Section 3}

Now let us pass to the statistical description of QTS and for simplicity we
start from the system of 2-order. Our main goal is to determine the
distribution function for nonequilibrium states of such system. Note that
under pure dynamical description the distribution function $f(x_{i},t)$ must
satisfy the continuity equation: $\frac{\partial f}{\partial t}+div(f\overset%
{.}{x}_{i})=0$. Because of attractors existence in QTS such function
necessarily must be singular. To get the smooth distribution function we
will use the following method. We introduce the interaction of QTS of
interest with ergometer - large exterior system which measures its energy
From the definition of ergometer it follows that during the measurement
process mean energy of QTS system remains the same. As we have showed in
\cite{3} such method results in Fokker-Planck equation for $f(x_{i},t)$ of
the form

\begin{equation}
\frac{\partial f}{\partial t}+\frac{\partial }{\partial x_{i}}(f\overset{.}{x%
}_{i})=\frac{\partial }{\partial x_{i}}(D_{ik}\frac{\partial f}{\partial
x_{k}}),  \label{15}
\end{equation}

where diffusion tensor $D_{ik}$ representing the influence of ergometer on
the system behavior must be chosen in the form:

\begin{equation}
D_{ik}=\varepsilon _{il}\varepsilon _{km}\frac{\partial H}{\partial x_{l}}%
\frac{\partial H}{\partial x_{m}},  \label{16}
\end{equation}

Here $H(x_{1},x_{2})$ is the energy of QTS. It is obviously that tensor $%
D_{ik}$ defined in such a way possesses the symmetry property: $%
D_{ik}=D_{ki} $ and the nonnegative property: $D_{11}>0,D_{22}>0$ and $%
D_{11}D_{22}-D_{12}^{2}>0$. Having the Fokker-Planck equation (15) in hand
we can search its stationary solutions corresponding to equations of motion:
$\overset{.}{x}_{i}=\varepsilon _{ik}\frac{\partial H}{\partial x_{k}}%
\left\{ S,H\right\} $. It is easy to show that all such solutions can be
represented in the form:

\begin{equation}
f(x_{1},x_{2})=A(H)e^{S(x_{1},x_{2})},  \label{17}
\end{equation}

where $S$ is the entropy of QTS and $A(H)$ is arbitrary function of its
energy. We define function $A(H)$ from the following physical reasons. First
of all let us pay attention to the fact that QTS entropy is defined up to
arbitrary function of energy because the transformation: $%
S(x_{1},x_{2})\rightarrow S(x_{1},x_{2})+C(H)$ does not change the form of
equations of motion. Besides we must require the distribution function (17)
transforms into well-known Gibbs distribution in the limiting case of
equilibrium state. It is easy to verify these conditions impose by the only
possible choice of distribution function in the form:

\begin{equation}
f(x_{1},x_{2})=Ne^{S(x_{1},x_{2})-S_{eq}(x_{1}^{0},x_{2}^{0})-\beta
H(x_{1},x_{2})},  \label{18}
\end{equation}

where $N$ is normalizing coefficient defined by the condition:

\begin{equation*}
\int f(x_{1},x_{2})dx_{1}dx_{2}=1,
\end{equation*}

and $S_{eq}(x_{1}^{0},x_{2}^{0})$ is the entropy in the equilibrium state
what is the function of the QTS energy and $\beta $ is the parameter having
sense of inverse temperature of system in the equilibrium state. The
relation (24) for distribution function of QTS is valid both in equilibrium
and nonequilibrium states and it is a main result of this section of the
paper. Note that relation of type (24) for distribution function can be
derived also on the base of information-theoretic Jaynes method \cite{4} but
we suppose these two methods are essentially different. The main difference
that in our method the "entropy" is actually the dynamic quantity defined by
equations of motion. So it can be used for describing both equilibrium and
nonequilibrium states of QTS. Also it is worth to note that for deriving the
distribution function (24) from Fokker-Planck equation the Markov approach
of system evolution is required. This fact is ignored in Jaynes method.

\section{Section 4}

In this section using the explicit form of distribution function (24) we
find the number of relations which are connecting the different
characteristics of QTS. First of all we point to simple relation between
relaxation time $\tau (E)$ and entropy of QTS in the equilibrium state. For
this purpose we write down the equation of motion for phase close to
equilibrium line $\varphi =\varphi _{0}(E)$. In the lowest approximation it
has a form: $\overset{.}{\varphi }=-\frac{[\varphi -\varphi _{0}(E)]}{\tau
(E)}$. Comparing this equation with (9) we derive:

\begin{equation}
\tau (E)=-\frac{1}{\left( \frac{\partial ^{2}S}{\partial \varphi ^{2}}%
\right) _{\varphi =\varphi _{0}(E)}}.  \label{19}
\end{equation}

It is obviously not far from equilibrium line QTS entropy may be written in
the following form up to terms of second order:

\begin{equation}
S(\varphi ,E)=S_{eq}(E)-\frac{\left( \varphi -\varphi _{0}(E)\right) ^{2}}{%
2\tau (E)}.  \label{20}
\end{equation}

Comparing (20) with (18) we derive the distribution function for NESS close
to equilibrium:

\begin{equation}
f(\varphi ,E)=Ne^{-\frac{\left( \varphi -\varphi _{0}(E)\right) ^{2}}{2\tau
(E)}-\beta E}.  \label{21}
\end{equation}

We denote by $\Delta \varphi $ the deviation of system phase from its
equilibrium value: $\Delta \varphi =\varphi -\varphi _{0}(E)$. Let us
evaluate the mean value $\left\langle \frac{(\Delta \varphi )^{2}}{\tau (E)}%
\right\rangle $ in NESS. From the definition of mean value we obtain

\begin{equation}
\left\langle \frac{(\Delta \varphi )^{2}}{\tau (E)}\right\rangle =\frac{\int
\frac{(\Delta \varphi )^{2}}{\tau (E)}e^{-\left( \frac{\Delta \varphi }{%
2\tau }\right) ^{2}-\beta E}d\varphi dE}{\int e^{-\left( \frac{\Delta
\varphi }{2\tau }\right) ^{2}-\beta E}d\varphi dE}=\frac{\int \sqrt{\tau (E)}%
e^{-\beta E}dE}{\int \sqrt{\tau (E)}e^{-\beta E}dE}=1.  \label{22}
\end{equation}

Relation $\left\langle \frac{(\Delta \varphi )^{2}}{\tau (E)}\right\rangle
_{NESS}=1$ is valid close to equilibrium state as one can see from its
derivation. Now we derive the relation which is valid not only close to
equilibrium. Let us consider the mean value $\left\langle \frac{\partial
(S-S_{eq})}{\partial E}\right\rangle _{NESS}$. Using the definition of mean
value and integrating by parts in numerator we obtain:

\begin{equation}
\left\langle \frac{\partial (S-S_{eq})}{\partial E}\right\rangle =\frac{\int
\frac{\partial (S-S_{eq})}{\partial E}e^{S-S_{eq}-\beta E}d\varphi dE}{\int
e^{S-S_{eq}-\beta E}d\varphi dE}=\beta .  \label{23}
\end{equation}

The obtained relation (29) connects mean value of $\frac{\partial (S-S_{eq})%
}{\partial E}$ in nonequilibrium stationary state with inverse temperature
of equilibrium state of the same mean energy. And finally let us set the
connection between the relaxation time $\tau (E)$ and inverse temperature $%
\beta $. For this purpose we consider the value

\begin{equation}
Z(\beta )=\int\limits_{0}^{\infty }\sqrt{\tau (E)}e^{-\beta E}dE.  \label{24}
\end{equation}

As follows from (22) the function $\sqrt{\tau (E)}e^{-\beta E}$ has the
sense of non-normalized distribution function of QTS close to equilibrium.
Now we find the mean energy of the system - $U$. By the definition we have:

\begin{equation}
U=\left\langle E\right\rangle =\frac{\int E\sqrt{\tau (E)}e^{-\beta E}dE}{%
\int \sqrt{\tau (E)}e^{-\beta E}dE}=-\frac{\partial }{\partial \beta }\ln
Z(\beta ).  \label{25}
\end{equation}

We suppose $\tau (E)$ to be monotonically increasing function of $E$ and
also suppose that the function $\sqrt{\tau (E)}e^{-\beta E}$ possesses the
sharp maximum on $E$.

Then we can find the value $Z(\beta )$ using the saddle point method: $%
Z(\beta )\approx \tau (U)e^{-\beta U}$, where mean energy $U$ is defined by
the condition of integrand function maximum in (30) $\frac{\frac{\partial
\tau }{\partial U}}{2\sqrt{\tau (U)}}-\beta \sqrt{\tau (U)}=0$. From that we
find the required connection between $\tau $ and $\beta $:

\begin{equation}
\beta (U)=\frac{1}{2}\frac{d}{dU}\ln \tau (U).  \label{26}
\end{equation}

Note though we investigated only the simplest case of QTS of 2-order
nevertheless all results obtained in Sec.3 and Sec.4 may be generalized with
the small modifications to the case of arbitrary $N$. For example for $N=3$
the diffusion tensor corresponding to the interaction with ergometer must be
chosen in the following form: $D_{ik}=\varepsilon _{ilm}\varepsilon _{kln}%
\frac{\partial H}{\partial x_{m}}\frac{\partial H}{\partial x_{n}}$. One can
see the form of stationary distribution function (24) derived as the
solution of Fokker-Planck equation (21) remains the same. We give also the
analog of (32) for QTS of $N$-order without derivation. In this case system
has the collection of relaxation times: $\tau _{1}(E),\tau _{2}(E),...\tau
_{N-1}(E)$. They are connected with the inverse temperature of the
equilibrium state by the relation:

\begin{equation}
\beta (E)=\frac{1}{2}\frac{d}{dE}\left[ \ln
\prod\limits_{i=1}^{N-1}\tau _{i}(E)\right] .  \label{27}
\end{equation}

\section{Conclusion}

Let us sum up the results of the paper. Our main goal was to demonstrate how
using Clausius conditions (1) one can successively construct classes of DS
of increasing complexity exactly satisfying to these conditions. We
demonstrated also that properties of such systems in the equilibrium
coincide with properties of usual thermodynamic systems. That's why the term
"quasithermodynamic system" was proposed. Besides using the interaction of
QTS of interest with ergometer we could give the statistical description of
nonequilibrium stationary states and obtain universal distribution function
for them. Note though we investigated the case of classical systems only the
proposed method can be extended to the quantum realm. Speaking more
accurately the equations of motion for QTS (4) admit the semiclassical
quantization, for example by using the method of paper \cite{5}. Detailed
investigation of quantum-mechanical variant of QTS description will be given
in future publications. Concerning to the possible application of obtained
results we must point out that they are highly diverse and may be applicable
not only in physics. We bring here only one such example concerning to the
chemical kinetics. It is known ( see e.g.\cite{6}) that for closed chemical
systems where the chemical reactions not higher then second-order on
concentrations of reagents $x_{i}\geqslant 0$ take place one can write down
the following system of kinetic equations:

\begin{equation}
\frac{dx_{i}}{dt}=\beta _{ik}x_{k}+\gamma _{ikl}x_{k}x_{l}.  \label{28}
\end{equation}

We suppose the absence of autocatalytic reactions in such system and
condition $\sum\limits_{i}\rho _{i}x_{i}=M$ where $\rho _{i}$ is molecular
mass of reagent $x_{i}$ is satisfied. Then it is easy to demonstrate that
all such systems are quasithermodynamic. Let us illustrate this statement on
the simplest case of two reagents $x_{1}$ and $x_{2}$ with coinciding
molecular masses: $\rho _{1}=\rho _{2}$.

In this case kinetic equations have the form:

\begin{equation}
\frac{dx_{1}}{dt}=-\alpha x_{1}+\beta x_{2}-\gamma x_{1}x_{2},\qquad \frac{%
dx_{2}}{dt}=\alpha x_{1}-\beta x_{2}+\gamma x_{1}x_{2}.  \label{29}
\end{equation}

One can easy see by the direct verification that using functions

\begin{equation}
H=x_{1}+x_{2},\qquad S=-\frac{\alpha x_{1}^{2}}{2}-\frac{\beta x_{2}^{2}}{2}-%
\frac{\gamma x_{1}^{3}}{6}-\frac{\gamma x_{1}^{2}x_{2}}{2}.  \label{29a}
\end{equation}

we can rewrite the system (35) in a standard form: $\overset{.}{x}%
_{i}=\varepsilon _{ik}\frac{\partial H}{\partial x_{k}}\left\{ S,H\right\} .$
This fact means that all results of the paper, in particular the possibility
of description of reagents concentrations fluctuations, become applicable to
the closed chemical systems. From this example we can see also that the
nature of functions $H$ and $S$ is defined exclusively by character of the
considered system that is by its equations of motion and may be essentially
different from their thermodynamic analogs.

\bigskip

\bigskip Author would acknowledge L.A. Pastur for the discussion of the
results of the paper and valuable comments.

\bigskip


\begin{thebibliography}{9}
\bibitem{1} R.Kubo Thermodynamics, (North-Holland, Amsterdam 1968)

\bibitem{2} R.I. McLachlan, G.R.W. Quispel, N. Robidoux Phys.Rev.Lett. 81,
2399 (1998)

\bibitem{3} E.D. Vol Int.J. of Theor.Phys. (will be published)

\bibitem{4} Jaynes E.T. Phys.Rev. 106, 620 (1957); 108,171 (1957)

\bibitem{5} E.D. Vol Phys.Rev A73, 062113 (2006)

\bibitem{6} N.G. Van Kampen Stochastic processes in physics and chemistry
(North-Holland, Amsterdam 1984)
\end{thebibliography}
\end{document}